\documentclass[aps,floats,floatfix,amssymb,prd,onecolumn,superscriptaddress,nofootinbib,notitlepage,11pt]{revtex4-1}

\usepackage[english]{babel}
\usepackage{latexsym,amssymb,amsmath,amsfonts,physics,url,mathrsfs}
\usepackage{graphicx}
\usepackage{comment}
\usepackage{hyperref}
\usepackage[table]{xcolor}
\usepackage{orcidlink}
\usepackage{tikz}

\newcommand{\be}{\begin{equation}}
\newcommand{\ee}{\end{equation}}

\newcommand{\tp}{\widetilde P_v}
\newcommand{\hm}{{\rm H}}
\hypersetup{
     colorlinks   = true,
     citecolor    = violet,
     urlcolor     = violet,
     linkcolor    = violet}

\def\md{\mathrm{d}}

\newcommand{\papertitle}{Non-linear Quasi-Normal Modes of the Schwarzschild \\ Black Hole from the Penrose Limit}

\begin{document}

\title[]{\papertitle}

\author{Alex Kehagias\orcidlink{ 0000-0001-6080-6215}}
\affiliation{Physics Division, National Technical University of Athens, Athens, 15780, Greece}

\author{Davide Perrone\orcidlink{0000-0003-4430-4914}}
\affiliation{Department of Theoretical Physics and Gravitational Wave Science Center,  \\
24 quai E. Ansermet, CH-1211 Geneva 4, Switzerland}

\author{Antonio Riotto\orcidlink{0000-0001-6948-0856}}
\affiliation{Department of Theoretical Physics and Gravitational Wave Science Center,  \\
24 quai E. Ansermet, CH-1211 Geneva 4, Switzerland}


\begin{abstract}
\noindent
The Penrose limit connects a plane wave geometry to the photon ring of a black hole, where the quasi-normal modes are located in the eikonal limit.
Utilizing this simplification, we analytically extract the quadratic-level non-linearities in the quasi-normal modes of a Schwarzschild black hole  for the $(\ell\times\ell)\to 2\ell$ channel.
We demonstrate that this result is independent of $\ell$ and further confirm it through symmetry arguments.

\end{abstract}

\maketitle
\section{Introduction}
\noindent
The identification of Quasi-Normal Modes (QNMs) in black hole ringdowns will facilitate consistency tests and independent confirmations of general relativity. 
To date, numerous studies have analyzed existing data \cite{LIGOScientific:2020tif,LIGOScientific:2021sio,Capano:2021etf, Finch:2022ynt, Isi:2022mhy, Cotesta:2022pci, Siegel:2023lxl} and made predictions for future ground- and space-based gravitational wave  detectors (see, for instance, Refs. \cite{Berti:2005ys, Ota:2019bzl, Bhagwat:2019dtm, Pitte:2024zbi}) based on linear ringdown QNM frequencies. 
However general relativity is inherently non-linear and recent studies have confirmed the existence of quadratic QNMs in relativistic numerical simulations of binary black hole mergers \cite{London:2014cma,Mitman:2022qdl, Cheung:2022rbm, Ma:2022wpv, Redondo-Yuste:2023seq, Cheung:2023vki, Zhu:2024rej}. 
Regarding their amplitudes, research has focused on deriving the quadratic-to-linear amplitude ratio using both numerical and analytical approaches \cite{Kehagias:2023ctr,Redondo-Yuste:2023seq,Cheung:2023vki,Perrone:2023jzq,Zhu:2024rej,Ma:2024qcv,Bourg:2024jme,Bucciotti:2024zyp,Khera:2024yrk,Kehagias:2024sgh,Bucciotti:2025rxa,bourg2025quadraticquasinormalmodesnull}. 
For instance, the amplitude $A_{2\times 2}$ of the non-linearly generated $\ell=4$ QNM, constructed out of two $\ell=2$ QNMs with amplitude $A_2$ for a Schwarzschild black hole is given by \cite{Redondo-Yuste:2023seq} 
\be
    {\cal R}_{2\times 2}=\left|\frac{A_{2\times 2}}{A_{2}^2}\right|\simeq 0.154.
\ee
This ratio exhibits only a mild dependence on black hole spin and decreases by a factor of approximately $0.5$ for highly spinning Kerr black holes. 
Such non-linear effects could potentially be detected in a few events using current ground-based observatories, with even greater prospects for observation through the LISA mission \cite{Yi:2024elj, Lagos:2024ekd}.

In this paper we take advantage of the fact that the QNMs, in the limit of large angular momenta (the so-called eikonal limit),  may be thought of as   degrees of freedom  populating the photon ring and slowly leaking out of it \cite{goebel1972comments}. 
The ``near-photon ring" physics can be  described by what is known as the Penrose limit which associates a plane wave to a region of spacetime near a null geodesic \cite{Reidel:1976}. 
Such plane wave  geometrically realizes the geometrical optics approximation valid in the large multipole limit.

The Penrose limit considerably simplifies the description of the QNMs \cite{Fransen:2023eqj,Giataganas:2024hil} and allows to recover analytically general results about the quadratic QNMs in the eikonal limit. 
We will show by a direct  calculation, supplemented by a simple symmetry argument, that the ratio ${\cal R}_{\ell\times \ell}$ does not depend on $\ell$ for $\ell\gg 1$ and we will calculate its asymptotic value for a Schwarzschild black hole.

The paper is organized as follows. 
In section \ref{sec:penrose_limit} we describe the Penrose limit and the corresponding first-order perturbations as well as its description in terms of the Weyl scalars. 
In section \ref{sec:second_order} we calculate ans solve the second-order equation for one the Weyl scalars, while in section \ref{sec:quadratic_qnm} we derive the quadratic QNMs for the gravitational waves for the channel $(\ell\times\ell)\to 2\ell$. 
We conclude in section \ref{sec:conclusions}. 
The paper is supplemented by three appendices, regarding the matching, the channel $(2\times\ell)\to \ell+2$ and the gauge invariance at second-order.

\section{The Penrose limit}
\label{sec:penrose_limit}
\subsection{Background}
\noindent
The Penrose limit \cite{Reidel:1976} associates to every space-time metric \( g_{ab} \), with line element \( {\rm d}s^2 \), and a null geodesic \( \gamma \) in that space-time, a (limiting) plane wave metric. 
The first step is to rewrite the metric in coordinates ``adapted" to \( \gamma \), which corresponds to an embedding of \( \gamma \) into a twist-free congruence of null geodesics, given by some coordinates \( V \), \( Z \), and \( \bar{Z} \) constant, with the remaining coordinate \( U \) playing the role of the affine parameter and \( \gamma(U) \) coinciding with the geodesic at \( V = Z = \bar{Z} = 0 \).  
The next step is to perform the change of coordinates  
\[
(U,V,Z,\bar{Z}) = (u,\lambda^2 v,\lambda z,\lambda \bar{z})
\]  
for some real \( \lambda \) and to take the limit  
\[
\lim_{\lambda\rightarrow 0} \lambda^{-2} {\rm d}s^2 = {\rm d}s^2_\gamma.
\]  
The resulting metric is the so-called Penrose limit of the initial space-time, which, recast in Brinkmann coordinates, has the plane-parallel (pp) wave form  

\begin{equation}
\label{eq:metric}
{\rm d}s^2_\gamma
= 2 {\rm d}u {\rm d}v + H(u,z,\bar{z}) {\rm d}u^2 - 2 {\rm d}z {\rm d}\bar{z}.
\end{equation}  
The coordinate \( u \) plays the role of the affine parameter along the geodesic, and the function \( H \) controls the geodesic deviation properties along the transverse coordinates \( z \) and \( \bar{z} \). For the  Schwarzschild black hole of mass $M$ and metric (where we have set the Newton constant and the light speed to unity)

\begin{eqnarray}
  {\rm d}s^2&=&f(r) {\rm d}t^2-f^{-1}(r){\rm d}r^2-r^2({\rm d}\theta^2+\sin^2\theta{\rm d}\phi^2),\nonumber\\
  f(r)&=&1-2M/r,
\end{eqnarray}
the Penrose limit  around the circular photon ring located at $r_0=3M$ and $\theta=\pi/2$ has the metric (\ref{eq:metric}) with
(see for instance Refs. \cite{Fransen:2023eqj,Giataganas:2024hil})
\begin{equation}
    H(z,\bar z)=-\frac{1}{3M^2}(z^2+\bar z^2),
\end{equation}
with $u=-t/3$ and $v=-t+3 \sqrt{3}M \phi$ in terms of Schwarzschild coordinates.

\subsection{Linear perturbations}
\noindent
For a generic pp-wave space-time such as the one described by the metric (\ref{eq:metric}) and satisfying the condition (where subscripts denote partial derivatives with respect to the corresponding variable, e.g.,$H_z = \partial_z H,$ and similarly for others)
\begin{equation}
 H_{z\bar{z}}=0,
\end{equation}  
any metric perturbation $h_{\mu\nu}$
that satisfies the linearized Einstein vacuum equations can be expressed in terms of a complex scalar field \( \Phi \), the Hertz potential,  which satisfies the Klein-Gordon equation 
\begin{equation}
\label{eq:hertz_potential}
\frac{1}{2} \Box\Phi=-\Phi_{z\bar z}-\frac{1}{2}H(z,\bar z) \Phi_{vv}+\Phi_{uv}=0.
\end{equation}
Indeed, having a solution of the scalar wave equation (\ref{eq:hertz_potential}), one  can construct solutions of  spin-two states by using the spin-raising operator. The latter is written in Brinkmann  coordinates   as \cite{Adamo:2017nia}
\begin{align}
    R^+=&\md u\, \partial_{\bar z}+\md z\, \partial_v,\nonumber \\
    R^-=&\md u\, \partial_{ z}+\md \bar z\, \partial_v.
\end{align}
For the  polarization vector
\begin{eqnarray}
\epsilon=\Big(0,0,\epsilon_z,
\epsilon_{ \bar z}\Big)=\Big(0,0,\epsilon_+,\epsilon_-\Big),
\end{eqnarray}
where 
\begin{eqnarray}
 \epsilon_+\epsilon_-=0,  \label{ee}
\end{eqnarray}
 the spin-two field  reads 
\begin{eqnarray}
    h_{\mu\nu}\md x^\mu \md x^\nu =h_{\mu\nu}^{++}\md x^\mu \md x^\nu +h_{\mu\nu}^{--}\md x^\mu \md x^\nu 
\end{eqnarray}
where
\begin{align}
    h_{\mu\nu}^{++}\md x^\mu \md x^\nu=&\epsilon_+ \epsilon_+ R^+\Big[R^+(\Phi)\Big], \nonumber \\
    h_{\mu\nu}^{--}\md x^\mu \md x^\nu=&\epsilon_- \epsilon_- R^-\Big[R^-(\bar \Phi)\Big].
    \label{hhmm}
\end{align}
 Normalizing the polarization vector such that $\epsilon_+^2=\epsilon_-^2=1$ and  in the radiation gauge (setting also the transverse-free and traceless conditions)
 \be
 \nabla^\mu h_{\mu\nu}=g^{\mu\nu}h_{\mu\nu}=0,\quad h_{v\mu}=0
 \ee 
 the  metric perturbed by the spin-two fields $h_{\mu\nu}$ is at first-order
\begin{eqnarray}
\label{eq:metricp1}
{\rm d} s^2&=&2{\rm d}u{\rm d}v+\left[H+\Phi_{\bar z\bar z}+\bar \Phi_{zz}\right] {\rm d}u^2
+
2\Phi_{v\bar z}{\rm d} u{\rm d}z+2\bar \Phi_{v z}{\rm d} u{\rm d}\bar z
\nonumber\\
&+& \Phi_{vv}{\rm d} z^2+
\bar \Phi_{vv}{\rm d} \bar z^2
-2{\rm d}z{\rm d}\bar z. \label{s0}
\end{eqnarray}

\subsection{The Weyl scalars}
\noindent
In the following it will prove convenient to work in terms of Weyl scalars. 
We  first need to find the appropriate null tetrad. 
The latter in the coordinates $(u,v,z,\bar z)$ is given  up to first-order by 
\begin{align}
    \ell_\mu=&a (1,0,0,0), \nonumber \\ n_\mu=&\frac{1}{a}\left[\frac{1}{2}(H+\Phi_{zz}+\bar{\Phi}_{\bar z\bar z}),1,0,0\right]\nonumber \\
    m_\mu=&\left(-\frac{1}{2}\Phi_{v\bar z},0,-\frac{1}{2}\Phi_{vv},1\right), \nonumber \\
    \bar m_\mu=&\left(-\frac{1}{2}\bar \Phi_{v z},0,1,-\frac{1}{2}\bar \Phi_{vv}\right), \label{mln}
\end{align}
so that the first-order metric (\ref{eq:metricp1}) can be also written as 
\begin{eqnarray}
{\rm d} s^2&=&2 \ell_{(\mu}n_{\nu)}-2 m_{(\mu}\bar m_{\nu)}. \label{eq:s0}
\end{eqnarray}
Note that there is an arbitrary normalization of the $\ell_\mu$ and $n_\mu$ tetrads, which eventually needs to be fixed. 
One way to do that is to consider the background metric at $z=\bar z=0$ given by 
\begin{equation}
    {\rm d} s^2=2{\rm d}u{\rm d}v
-2{\rm d}z{\rm d}\bar z.
\end{equation}
The corresponding tetrads (\ref{mln}) are then given by (here the superscript indicates the order of the expansion)
\begin{align}
    \ell^{(0)}_\mu=&a (1,0,0,0), \nonumber \\ n^{(0)}_\mu=&\frac{1}{a}\left(0,1,0,0\right)\nonumber \\
    m^{(0)}_\mu=&\left(0,0,0,1\right), \nonumber \\
    \bar m^{(0)}_\mu=&\left(0,0,1,0\right).
    \label{mln0}
\end{align}
Now,  consider a perturbations of the form
\begin{equation}
    {\rm d} s^2=2{\rm d}u{\rm d}v +(h_+-ih_\times){\rm d}z^2+(h_++ih_\times){\rm d}\bar z^2-2{\rm d}z{\rm d}\bar z. \label{s1}
\end{equation}
A comparison with  Eq. (\ref{eq:s0}) shows that 
\begin{eqnarray}
\label{comp}
    \Phi_{vv}= (h_+-ih_\times), \qquad \bar \Phi_{vv}=(h_++ih_\times). 
\end{eqnarray}
Then by defining the Weyl scalars 
\begin{align}
    \Psi_0=&-C_{\mu\nu\rho\sigma}\,\ell^\mu m^\nu \ell^\rho m^\sigma,\nonumber \\
    \Psi_1=&-C_{\mu\nu\rho\sigma}\,\ell^\mu n^\nu \ell^\rho m^\sigma,\nonumber \\
    \Psi_2=&-C_{\mu\nu\rho\sigma}\,\ell^\mu m^\nu \bar m^\rho n^\sigma ,\nonumber \\
    \Psi_3=&-C_{\mu\nu\rho\sigma}\,\ell^\mu n^\nu \bar m^\rho n^\sigma, \nonumber \\
    \Psi_4=& -C_{\mu\nu\rho\sigma}\,n^\mu \bar m^\nu n^\rho \bar m^\sigma, \label{weyl}
\end{align}
we can find that two of the Weyl scalars we will be interested below, $\Psi_0$ and $\Psi_4$, are 
\begin{align}
    \Psi_0=&\frac{1}{2}a^2(h_+-ih_\times)_{vv}, \label{psi0} \\
    \Psi_4=&\frac{1}{2a^2}(h_++ih_\times)_{vv}.
    \label{psi4}
\end{align}
The corresponding non-zero spin coefficients in  the basis (\ref{mln}) up to first-order  are 
\begin{align}
    \begin{split}
        &\tau^{(1)}=-\frac{1}{2}\Phi_{vv\bar z},\\
       & \rho=0, \\
       &\sigma^{(1)}=\frac{1}{2}a\Phi_{vvv}, \\
       &\kappa=0,\\
     &  \gamma^{(1)}=\frac{1}{2a}\Phi_{v\bar z\bar z},\\
     &\lambda^{(1)}=\frac{1}{a}\left(\bar \Phi_{vz\bar z}+\frac{1}{4}H \bar \Phi_{vvv}-\frac{1}{2}\bar \Phi_{uvv}\right),\\
     &\mu^{(1)}=\frac{1}{2a}\left(\Phi_{v\bar z\bar z}+\bar \Phi_{vzz}\right),
    \end{split}\qquad
    \begin{split}
       & \alpha=0,\\
        &\beta^{(1)}=-\frac{1}{2}\Phi_{vv\bar z},\\
        &\epsilon=0,\\
        & \nu^{(0)}=-\frac{1}{2a^2}H_{\bar z}\\
        &\nu^{(1)}=-\frac{1}{4a^2}\left(2\Phi_{\bar z\bar z\bar z}+2\bar \Phi_{zz\bar z}+
        H_z\bar \Phi_{vv}\right)\\
        &\pi^{(1)}=-\frac{1}{2}\bar \Phi_{vvz},\\
        &
    \end{split}
    \label{spin}
\end{align}
whereas the non-zero Weyl scalars turn out to be
\begin{align}
&\Psi_0^{(1)}=\frac{1}{2}a^2\Phi_{vvvv}\qquad 
\Psi_1^{(1)}=-\frac{1}{2}a \Phi_{vvv\bar z}, \qquad
        \Psi_2^{(1)}=\frac{1}{2}\Phi_{vv\bar z\bar z},\nonumber \\
        &
        \Psi_3^{(1)}=-\frac{1}{2a}\Phi_{v\bar z\bar z\bar z},\qquad  \Psi_4^{(0)}=\frac{1}{2a^2}H_{\bar z\bar z}
        \label{w1}
\end{align}
and 
\begin{align}
     \Psi_4^{(1)}=&\frac{1}{a^2}\bigg(\frac{1}{2}\Phi_{\bar z\bar z\bar z \bar z}+\frac{1}{2}\bar \Phi_{zz\bar z\bar z }+\frac{1}{4}H_z \bar \Phi_{vv\bar z} +\frac{1}{4}H_{\bar z} \bar \Phi_{vv z}+\frac{1}{2} H\bar \Phi_{vv z\bar z}\nonumber \\
    +&\frac{1}{2}H_{z\bar z} \bar\Phi_{vv}+\frac{1}{8} H^2 \bar \Phi_{vvvv}-\bar \Phi_{uvz\bar z}-\frac{1}{2}H \bar \Phi_{uvvv}+\frac{1}{2}\bar \Phi_{uuvv}\bigg).  \label{w2}
\end{align}

\section{Second-order equation for the Weyl scalar $\Psi_0$}
\label{sec:second_order}
\noindent
For Petrov-type D spacetimes like the Kerr black hole, the first-order $\Psi_4^{(1)}$ is completely determined by the zeroth-order  background (Teukolsky equation) and similarly, the second-order $\Psi_4^{(2)}$ is determined purely by first-order quantities. 
For Petrov-type N, like the pp-wave background we are considering here, this is not  true any longer. 
In fact, the first-order  $\Psi_4^{(1)}$ is determined now by the other first-order Weyl scalars and similarly the second-order  $\Psi_4^{(2)}$ is determined by the second-order quantities. 
In other words, to determine  $\Psi_4^{(2)}$, we need all the second-order perturbations of the metric. 
Therefore, for Petrov-type N spacetimes, it is more convenient to consider $\Psi_0$ instead. 
In order to determine the equation for $\Psi_0$, we will need the following two Bianchi identities for Ricci-flat spacetimes
\begin{align}
    &-(\bar \delta+\pi-4\alpha) \Psi_0
    +(D-4\rho -2\epsilon)\Psi_1+3\kappa \Psi_2=0, \label{b1}\\
    &-(\Delta +\mu -4\gamma)\Psi_0+(\delta -4\tau-2\beta)\Psi_1+3\sigma 
    \Psi_2=0,
\label{b2}
\end{align}
where
\begin{eqnarray}
D=\ell^\mu \nabla_\mu, \qquad \Delta=n^\mu \nabla_\mu, \qquad     \delta=m^\mu\nabla_\mu,\qquad \bar \delta=\bar m^\mu\nabla_\mu.
\end{eqnarray}
Let us now act on Eq. (\ref{b1}) with $\delta^{(0)}$ and on Eq. (\ref{b2}) with $D^{(0)}$ and substract them. The result is the equation 
\begin{align}
   -& \bigg[\delta^{(0)} (\bar \delta+\pi-4\alpha)-D^{(0)}(\Delta +\mu -4\gamma) \bigg]\Psi_0\nonumber \\
  +&\bigg[ \delta^{(0)} (D-4\rho -2\epsilon)
  -D^{(0)}(\delta -4\tau-2\beta)
  \bigg]  \Psi_1\nonumber \\
  +&3\left(\delta^{(0)}\kappa-D^{(0)}\sigma\right) \Psi_2=0. \label{meq}
\end{align}
By the perturbative expansion of the Weyl scalars and the spin coefficients, we find that up to second-order Eq. (\ref{meq}) is written as 
\allowdisplaybreaks
\begin{align}
    -& \bigg[\delta^{(0)} (\bar \delta+\pi-4\alpha)^{(0)}-D^{(0)}(\Delta +\mu -4\gamma)^{(0)} \bigg]\Psi_0^{(1)}\nonumber \\
    -& \bigg[\delta^{(0)} (\bar \delta+\pi-4\alpha)^{(1)}-D^{(0)}(\Delta +\mu -4\gamma)^{(1)} \bigg]\Psi_0^{(1)}\nonumber \\
    -& \bigg[\delta^{(0)} (\bar \delta+\pi-4\alpha)^{(0)}-D^{(0)}(\Delta +\mu -4\gamma)^{(0)} \bigg]\Psi_0^{(2)}\nonumber \\
    +&\bigg[ \delta^{(0)} (D-4\rho -2\epsilon)^{(0)}
  -D^{(0)}(\delta -4\tau-2\beta)^{(0)}
  \bigg]  \Psi_1^{(1)}\nonumber \\
    +&\bigg[ \delta^{(0)} (D-4\rho -2\epsilon)^{(1)}
  -D^{(0)}(\delta -4\tau-2\beta)^{(1)}
  \bigg]  \Psi_1^{(1)}\nonumber \\
  -&3D^{(0)}\left(\sigma^{(1)} \Psi_2^{(1)}\right)=0.
  \label{tt}
\end{align}
Note that 
\begin{align}
   & \bigg[ \delta^{(0)} (D-4\rho -2\epsilon)^{(0)}
  -D^{(0)}(\delta -4\tau-2\beta)^{(0)}
  \bigg]  \Psi_1^{(2)}\nonumber \\
 =& \left(\delta^{(0)}D^{(0)}-D^{(0)}\delta^{(0)}\right)\Psi_1^{(2)}=0,
\end{align}
due to the commutation relation 
\begin{eqnarray}
    [\delta,D]=(\bar \alpha +\beta -\bar \pi) D+\kappa \Delta -(\bar \rho+\epsilon-\bar \epsilon)\delta -\sigma \bar \delta,
\end{eqnarray}
so  that
\begin{eqnarray}
    \left[\delta^{(0)},D^{(0)}\right]  =0.
\end{eqnarray}
 Eq. (\ref{tt}) can explicitly be written by using Eqs. (\ref{spin}), (\ref{w1}) and (\ref{w2}) and the following explicit form of the differential operators $\delta, D$ and $\Delta$
\begin{align}
    \delta^{(0)}=&-\partial_z, \qquad \delta^{(1)}=-\frac{1}{2}\Phi_{uu}\partial_{\bar z}, \nonumber \\
    D^{(0)}=&a\partial_v, \qquad \Delta^{(0)}
    =\frac{1}{a}\left( \partial_u-\frac{1}{2}H\partial_v\right), \nonumber\\
    \Delta^{(1)}=&-\frac{1}{2a}(\Phi_{\bar z\bar z}+\bar \Phi_{zz})\partial_v+\frac{1}{a}(\Phi_{v\bar z}\partial_{\bar z}+\bar{\Phi}_{vz}\partial_z).
\end{align}
Then it turns out from Eq.(\ref{tt}) that the second-order $\Psi_0^{(2)}$ satisfies the equation 
\begin{equation}
    T_2\Psi_0^{(2)}=S_2^{(2)}
\end{equation}
where the differential operator $T_2 $ is given by
\begin{align}
    T_2=-\delta^{(0)} (\bar \delta+\pi-4\alpha)^{(0)}+D^{(0)}(\Delta +\mu -4\gamma)^{(0)} , 
    \label{TT}
\end{align}
and the source $S^{(2)}_{\Psi_0}$ is 
\begin{align}
S^{(2)}_{\Psi_0}=&\bigg[\delta^{(0)} (\bar \delta+\pi-4\alpha)^{(1)}-D^{(0)}(\Delta +\mu -4\gamma)^{(1)} \bigg]\Psi_0^{(1)}\nonumber \\
    -&\bigg[ \delta^{(0)} (D-4\rho -2\epsilon)^{(1)}
  -D^{(0)}(\delta -4\tau-2\beta)^{(1)}
  \bigg]  \Psi_1^{(1)}\nonumber \\
+&3D^{(0)}\left(\sigma^{(1)} \Psi_2^{(1)}\right)
    \label{SS}
\end{align}
We can now use Eqs. (\ref{spin}), (\ref{w1}) and (\ref{w2}) in (\ref{TT}) from where we find that 
\begin{eqnarray}
   T_2\Psi_0^{(2)}= \frac{1}{2}\Box \Psi_0^{(2)}.
\end{eqnarray}
and calculate the source, which turns out to be
  
\begin{align}
\label{eq:source}
S^{(2)}_{\Psi_0}=&a^2\bigg[-\frac{3}{2} \Phi_{vvv\bar z}^2+\Phi_{vvv}\Phi_{vvv\bar z \bar z}
    +\frac{3}{2}
\Phi_{vv \bar z\bar z} \Phi_{vvvv}
\nonumber \\
-& 2 \Phi_{vv\bar z}
\Phi_{vvvv\bar z}+\frac{1}{4}  \Phi_{vv} \Phi_{vvvv\bar z\bar z}+
\frac{1}{4}\bar \Phi_{vv} \Phi_{vvvvzz}\nonumber \\
+&\Phi_{v \bar z \bar z}\Phi_{vvvvv}-\frac{1}{2}\Phi_{v\bar z}\Phi_{vvvvvv\bar z}
- \frac{1}{2}\bar \Phi_{vz} \Phi_{vvvvvz}
\nonumber \\
+&\frac{1}{4}\Phi_{\bar z\bar z} \Phi_{vvvvvv}+\frac{1}{4}\bar \Phi_{zz}\Phi_{vvvvvv}
\bigg]. 
\end{align}
For the QNMs, we first look for linear solutions to the Eq. (\ref{eq:hertz_potential}) and  demand an outgoing boundary condition for the unstable direction $r-r_0=(z+\bar z)/\sqrt{6}$, and, additionally, we impose a decaying boundary condition for the stable direction $\theta-\pi/2=i(\bar z-z)/r_0\sqrt{2}$. 
When  dealing only with the fundamental mode, they yield the scalar function to be of the form \cite{Fransen:2023eqj,Giataganas:2024hil}
\begin{eqnarray}
\label{source2}
    \Phi(u,v,z,\bar z)=A e^{-i P_u u+i P_v v +\frac{3}{4} \ell \omega^2(1+i)(z^2+2i z \bar z+\bar z^2)}, 
\end{eqnarray}
where 
\begin{equation}
    P_u=\frac{3}{2}\omega (i-1),\qquad 
    P_v=\ell \omega\equiv \omega_\ell, \qquad \omega=\frac{1}{3\sqrt{3} M}.
    \label{pupv}
\end{equation}
When substituting this solution the source in Eq. (\ref{eq:source}) becomes
\begin{align}
    S_{\Psi_0}^{(2)}=& -6(1+i)a^2 \ell^7 \omega^8
    \Phi^2 -\frac{3}{4} a^2 \ell^7 \omega^8 \Big[ 1+3 \ell \omega^2(z-\bar z)^2\Big]
 \Phi \bar \Phi.
\end{align}
Then, it is straightforward to verify that the solution to the equation 
\begin{eqnarray}
    \frac{1}{2}\Box \Psi_0^{(2)}=S_{\Psi_0}^{(2)},
\end{eqnarray}
 is the sum of two functions, one oscillating and decaying, the other only decaying.  The oscillating part   satisfies 
 \begin{eqnarray}
     \frac{1}{2}\Box \Psi_{0\,{\rm osc}}^{(2)}=-6(1+i)a^2 \ell^7 \omega^8
    \Phi^2,
 \end{eqnarray}
 and its solution is
\begin{eqnarray}
    \Psi_{0\,{\rm osc}}^{(2)}=2i a^2P_v^6\Phi^2 . \label{psi02}
\end{eqnarray}
Similarly, the  the non-oscillating (decaying) part 
satisfies 
\begin{eqnarray}
     \frac{1}{2}\Box \Psi_{0\,{\rm dec}}^{(2)}=-\frac{3}{4} a^2 \ell^7 \omega^8 \Big[ 1+3 \ell \omega^2(z-\bar z)^2\Big],
 \end{eqnarray}
and it is given by 
\begin{eqnarray}
    \Psi_{0\, { \rm dec}}^{(2)}=-\frac{1}{4} a^2\ell^6 \omega^6\Phi \bar \Phi. \label{dec}
\end{eqnarray}
We will not consider further the decaying solution (\ref{dec}) since it describes other effects (e.g. memory) we are not interested in.

\section{Quadratic QNMs for the channel  $(\ell\times\ell)\to 2\ell$}
\label{sec:quadratic_qnm}
\noindent
We are now in the position to calculate the non-linear QNMs of the gravitational waves  on the photon ring. 
This is not yet the full answer as we will need to propagate it awy from the photon ring at large distances. 
Using  Eqs. (\ref{comp}) and (\ref{psi0}) we find 
\begin{eqnarray}
    \Psi_0=\frac{1}{2}a^2 h_{vv}=-\frac{1}{2} a^2 P_v^2 h,
\end{eqnarray}
and
\begin{eqnarray}
    h=\frac{2\Psi_0}{a^2 P_v^2},
\end{eqnarray}
where we have left the arbitrary normalization $a$, to check that the non-linear ratio  is independent from it. 
From Eqs. (\ref{psi02}) and (\ref{w1}) we obtain 
\begin{align}
    h^{(2)}=&\frac{2\Psi_{0\,{\rm osc}}^{(2)}}{a^2 (2P_v)^2}=P_v^4 \Phi^2 , \label{h2}\\
    h^{(1)}=&\frac{2\Psi_0^{(1)}}{a^2 P_v^2}=P_v^2 \Phi. \label{h1}
\end{align} 
We finally obtain
\begin{equation}
    \label{eq:nonlinear_relation}
\frac{h^{(2)}}{\big(h^{(1)}\big)^2}=1.
\end{equation}
 Notice that we have calculated this ratio in the radiation gauge augmented with the traceless and transverse-free condition, which is the same gauge where the non-linearities are extracted numerically \cite{Loutrel:2020wbw,Ripley:2020xby}.
 In this way, the issues about the non gauge-invariance of the Weyl scalars at second-order (see, Appendix \ref{app:gauge_invariance} and Ref. \cite{Kehagias:2024sgh}) are avoided.

We have  found that in the Penrose limit of large multipoles the non-linear QNM ratio does not depend on the multipole itself for the process $(\ell\times\ell)\to 2\ell$. 
In the followig subsection we offer a simple symmetry argument to explain such a result to explain why $h^{(2)}/( h^{(1)}\big)^2$ is independent from the multipoles.

\subsection{Symmetry argument}
\label{sec:symmetry_argument}
\noindent
We  perform the following change of coordinates
\be
\label{change}
 u'=u,\,\,
 v'=\lambda^2\, v,\,\,
 z'=\lambda\, z,\,\,
 \bar z'=\lambda\,\bar z,
\ee
for which  
\begin{equation}
    \Box'=\lambda^{-2}\,\Box,
\end{equation}
since $H(z',\bar z')=\lambda^2\, H(z,\bar z)$. Therefore, if $\Phi$ is the solution of the equation $\Box\Phi=0$, then  
\begin{equation}
 \Phi(u',v',z',\bar z')=\lambda^\Delta\Phi(u,v,z,\bar z)  
\end{equation}
is still a solution of the equation $\Box\Phi=0$ for any $\Delta$. From Eq. (\ref{w1}) we read off that 
$\Psi_0^{(1)}$ has weight $(\Delta-8)$ and  the second-order source has weight
$(2\Delta-14)$. Therefore, assigning a weight $\beta$ to $\Psi_0^{(2)}$, we have
\begin{equation}
 -2+\beta=2\Delta-14,   
\end{equation}
from the equation of motion, from which we read off that $\beta=(2\Delta-12)$. This means that $
{\cal R}_{\Psi_0}$
has weight 
\begin{equation}
    (2\Delta-12-2\Delta+8)=-4.
\end{equation}
We  now can choose $\lambda=\ell^{1/2}$ to cancel the dependence on $\ell$ from the function $\Phi$ so that  $
{\cal R}_{\Psi_0}$ scales like $1/\ell^2$. Since $h^{(2)}/( h^{(1)}\big)^2$ scales like $\ell^2 {\cal R}_{\Psi_0}$, we  automatically deduce that the $h^{(2)}/( h^{(1)}\big)^2$  may not   depend on $\ell$ in the eikonal limit.

\subsection{An alternative derivation}
\noindent
It is possible to calculate directly the second-order metric perturbations 
$h_{\mu\nu}^{(2)}$ directly from second-order perturbations of the Einstein equations. 
For this, let us write the traceless second-order metric in the radiation gauge as 
\begin{eqnarray}
{\rm d} s^2&=&2{\rm d}u{\rm d}v+\left[H+\Phi_{\bar z\bar z}+\bar \Phi_{zz}+h_{uu}^{(2)}\right] {\rm d}u^2
+
\Big(2\Phi_{v\bar z}+h_{uz}^{(2)}\Big){\rm d} u{\rm d}z+\Big(2\bar \Phi_{v z}+h_{u\bar z}^{(2)}\Big){\rm d} u{\rm d}\bar z
\nonumber\\
&+& \Big(\Phi_{vv}+h_{zz}^{(2)}\Big){\rm d} z^2+
\Big(\bar \Phi_{vv}+h_{\bar z\bar z}^{(2)}\Big){\rm d} \bar z^2
-2{\rm d}z{\rm d}\bar z. \label{s1}
\end{eqnarray}
In addition, the transversality condition gives the relations
\begin{eqnarray}
\label{gcon}
    \partial_{\bar z}h_{uz}^{(2)}
    +\partial_{ z}h_{u\bar z}^{(2)}
    -\partial_{v}h_{uu}^{(2)}=0, \qquad
\partial_{v}h_{uz}^{(2)}=
\partial_{\bar z}h_{zz}^{(2)}, \qquad
\partial_{v}h_{u\bar z}^{(2)}=
\partial_{ z}h_{\bar z\bar z}^{(2)}.
\end{eqnarray}
Then, the second-order vacuum Einstein equations are 
\begin{eqnarray}
    R_{\mu\nu}^{(2)}=0,
\end{eqnarray}
and in particular it turns out that 
the equations 
\begin{eqnarray}
    R_{zz}^{(2)}=R_{\bar z\bar z}^{(2)}=0,
\end{eqnarray}
are explicitly given by 
\begin{eqnarray}
   && \partial_{v}\partial_z h_{uz}^{(2)}-\frac{9}{2}\omega^2(z^2+\bar z^2) \partial_v^2h_{zz}^{(2)}-\partial_u\partial_v h_{zz}^{(2)}+3(1+i)\ell^5 \omega^6\Phi^2\nonumber \\
   &&+\frac{3}{4} \ell^5 \omega^6
   \bigg\{2-2i+3\ell \omega^2\left[(z-\bar z)^2-2i(z^2-\bar z^2)\right] \bigg\}\Phi \bar \Phi=0,
   \label{eqq}
\end{eqnarray}
and its complex conjugate. Using the gauge condition (\ref{gcon}), we find that (\ref{eqq}) is written as 
\begin{eqnarray}
    \frac{1}{2}\Box h_{zz}^{(2)}=3(1+i)\ell^5 \omega^6\Phi^2+ \frac{3}{4} \ell^5 \omega^6
   \bigg\{2-2i+3\ell \omega^2\left[(z-\bar z)^2-2 z \bar z-2i(z^2-\bar z^2)\right] \bigg\}\Phi \bar \Phi.
\end{eqnarray}
The solution of the above equation sourced by the oscillating part of the right-hand source proportional to $\Phi^2$ is given by 
\begin{eqnarray}
    h_{zz}^{(2)}=h_+^{(2)}-ih_\times^{(2)}=i \ell^4 \omega^4 \Phi^2. 
\end{eqnarray}
By recalling that 
\begin{eqnarray}
   h_{zz}^{(1)} =h_+^{(1)}-ih_\times^{(1)}=\Phi_{vv}=-\ell^2\omega^2 \Phi, 
\end{eqnarray}
 we find that 
 \begin{eqnarray}
     \frac{ h^{(2)}_{2\ell}}{\big( h^{(1)}_\ell\big)^2}=1, 
 \end{eqnarray}
which confirms the result (\ref{eq:nonlinear_relation}).

\subsection{From the photon ring to large distances}
\noindent
The last step to find the ratio ${\cal R}_{\ell\times \ell}$ is to
match the Penrose limit solutions at first- and second-order  to the one at large distances away from the photon ring. This procedure is done in detail in Appendix \ref{app:matching} and we report here only the final result,

\begin{equation}
    \mathcal{R}_{\ell\times \ell}= \frac{1}{2}\frac{1}{\sqrt{\gamma_\ell}} \cdot\frac{c_{\ell}^2}{c_{2\ell}} \frac{ h^{(2)}_{2\ell}}{\big( h^{(1)}_\ell\big)^2}=\frac{1}{2}\frac{1}{\sqrt{\gamma_\ell}}\cdot\frac{c_{\ell}^2}{c_{2\ell}},
\end{equation}
where

\begin{equation}
\label{eq:cl}
c_\ell={}_{-2}Y_{\ell\ell}
\left(\frac{\pi}{2},0\right)=\frac{(-1)^\ell}{2^\ell }\sqrt{\frac{2\ell+1}{4\pi}\frac{ (2\ell)!}{ (\ell+2)! (\ell-2)!}}
\end{equation}
and 

\be
\gamma_\ell = e^{-i \pi/4} \left(2 Q_0''\right)^{1/4},
\ee
with \cite{Schutz:1985km}

\begin{equation}
 Q_0''=-\left(1-\frac{r_s}{r_0}\right)\frac{\ell^2}{r_0^4} \left(6 
 -20 \frac{r_s}{r} +15 \frac{r_s^2}{r^2}\right) 
\end{equation}
Here $\lambda=\ell(\ell+1)$, and $r_0=3r_s/2$ is the point of the maximum of the effective potential. In the eikonal limit $|{\cal R}_{\ell\times\ell}|$ goes to a plateau whose amplitude is of the order of 0.24. 
Indeed, for large $\ell$ we have 
\begin{eqnarray}
    c_\ell\approx \frac{e^{-2/\ell}\ell^{1/4}}{\sqrt{2}\pi^{3/4}}, \qquad
    \gamma_{\ell}\approx \frac{\sqrt{2}\ell^{1/2}}{3\sqrt{3}} e^{-i\pi/4},
\end{eqnarray}
and therefore, $|{\cal R}_{\ell\times\ell}|$ asymptotes to 
\begin{eqnarray}
    |{\cal R}_{\ell\times\ell}|\approx
    \frac{1}{4}\left(\frac{3}{\pi}\right)^{3/4}\approx 0.24.
\end{eqnarray}

\section{Conclusions}
\label{sec:conclusions}
\noindent
In this paper we have calculated the
level of non-linearities in the QNMs for the Schwarzschild black hole in the eikonal limit and at the quadratic level, focussing on the $(\ell\times \ell)\to 2\ell$ channel. 
We have adopted the Penrose limit which allows to zoom in onto the photon ring where the QNMs are generated. 
Our findings can be extended in several directions. 
First of all, it will be interesting to consider other channels, e.g. different multipoles as initial states. 
Maybe more interestingly, it would be relevant to extend our calculations to the Kerr black holes and confirm that the non-linearities do not scale much with the black hole spin. 
We leave such investigations for the next future \cite{inprep}.

\section*{Acknowledgements}
\noindent
A.K. acknowledges support from the Swiss National Science Foundation (project number
IZSEZ0 229414). A.R. acknowledges support from the Swiss National
Science Foundation (project number CRSII5 213497) and from the Boninchi Foundation for the
project “PBHs in the Era of GW Astronomy”. The work of D. P. is supported by the Swiss National Science Foundation under grants no. 200021- 205016 and PP00P2-206149.

\appendix
\setcounter{equation}{0}
\setcounter{section}{0}
\setcounter{table}{0}
\makeatletter
\renewcommand{\theequation}{A\arabic{equation}}

\section{Matching}
\label{app:matching}
\noindent
In this appendix we will study the matching needed to connect the result in the Penrose limit to the asymptotic solutions. 
Note that, in order to be consistent with the existing literature on the WKB approximation for the QNMs, we will change the signature used so far from \textit{mostly minus} to \textit{mostly plus}, so that our result can be easily checked.

We will perform the matching of the Hertz potential $\Phi$ from the photon ring to large distances in three steps:
\begin{itemize}
    \item First we match the angular part with the associated spherical harmonic at $\theta=\pi/2$, which is just proportional to a pure phase $e^{i \ell \phi}$;
    \item then, by noticing that the equation of motion  for the radial part are the same to that of an harmonic oscillator, we will match within the WKB approximation of an harmonic oscillator at $r=r_0$;
    \item finally we will match this WKB solution near the lightring with the WKB for the asymptotic behavior.
\end{itemize}
This procedure will give us a solution of the form
\begin{equation}
    \Phi_{\ell,m,n} =b_\Phi \cdot {}_{-2}Y_{\ell,m} \,e^{-i\omega_{\ell,m,n}(t-r)}
\end{equation}
at large distances away from the photon ring. 
The first point can be implemented by  recalling that  
\begin{equation}
    {}_{-2}Y_{\ell\ell}\left(\frac{\pi}{2},\phi\right)= c_{\ell}\,e^{i\ell \phi},
\end{equation}
where $c_\ell$ is given in Eq. (\ref{eq:cl}). 
The matching to the previous solution requires a constant in front of the asymptotic solution such that 
\begin{equation}
     e^{i\ell \phi} \to \frac{1}{c_{\ell}}{}_{-2}Y_{\ell\ell}\left(\frac{\pi}{2},\phi\right),
\end{equation}
which can then be extended for every value of $\theta$.

The second part is more complex and requires the study of the solution of the equation, (written in real coordinates $x_1=(z+\bar z)/\sqrt{2}$ and $x_2=(z-\bar z)/(i\sqrt{2})$),
 
\begin{equation}
    (2\partial_u\partial_v - \alpha^2(x_1^2-x_2^2)\partial_v^2 + \partial_1^2+\partial_2^2)\Phi = S,
\end{equation}
where   $\alpha^2=1/3M^2$ and    $S\sim (\Phi^{(1)})^2$.
To familiarize with the form of the equation we recap the homogeneous solution for the equation by writing the ansatz
\begin{equation}
\label{eq:Phi_solution}
    \Phi=b_{\Phi}\, e^{i P_v v+ i P_u u} \phi_1(x_1)\phi_2(x_2),
\end{equation}
and substituting back in the homogeneous equation we get
\begin{equation}
    -2P_u P_v \phi_1(x_1)\phi_2(x_2) + \alpha^2 P_v^2 (x_1^2-x_2^2)\phi_1(x_1)\phi_2(x_2)  + \phi_1(x_1) \phi_2''(x_2) + \phi_2(x_2) \phi_1''(x_1)=0.
\end{equation}
Splitting the two equations for $\phi_1$ and $\phi_2$ we get
\begin{align}
    \phi_1''(x_1) +\alpha^2 P_v^2\,x_1^2 \phi_1(x_1) = (P_uP_v + \Delta ) \phi_1(x_1),\\
    \phi_2''(x_2) - \alpha^2 P_v^2\,x_2^2 \phi_2(x_2) = (P_uP_v - \Delta) \phi_2(x_2),
\end{align}
with $\Delta$ being the splitting constant. We now perform the following change of variables
\begin{equation}
\label{eq:variable_change_to_ho}
    y_1 = \gamma\,  x_1, \quad y_2=\rho\, x_2,
\end{equation}
\begin{equation}
    \gamma= e^{-i\pi/4}\sqrt{2\alpha P_v},\quad \rho=\sqrt{2\alpha P_v},
\end{equation}
\begin{equation}
\label{eq:splitting_constants}
    i\frac{P_u P_v +\Delta}{2\alpha P_v} = - \left( n_1 + \frac 12\right), \quad \frac{P_u P_v - \Delta}{2\alpha P_v} = - \left( n_2 + \frac 12\right).
\end{equation}
This brings both equations in the form
\begin{equation}
\label{eq:harmonic_oscillator}
    \phi_i'' -\frac 14 y_i^2 \phi_i = -\left( \frac 12 +n_i\right)\phi_i, \qquad i=1,2,
\end{equation}
or 
\begin{equation}
    \hm_i\phi_i = E_i \phi_i,
    \end{equation}
    where 
    \begin{equation}
    \hm_i= \partial_{y_i}^2 - \frac 14 y_i^2, \qquad E_i = -\left(n_i + \frac 12\right),
\end{equation}
which is the 
 equation for an harmonic oscillator.
Furthermore to ensure both boundary conditions 
\begin{align}
    &e^{i\alpha P_v x_1^2/2},~\,\qquad  x_1\to \pm \infty, \\
   & e^{-\alpha P_v x_2^2/2},\qquad  x_2\to \pm \infty,
\end{align}
it is necessary to choose the decaying solution for both cases in the variables $y_i$, because
\begin{equation}
    e^{-y_1^2/4}\to e^{-\gamma^2x_1^2/4}= e^{i \alpha P_v x_1^2/2}, \qquad e^{-y_2^2/4}\to e^{-\rho^2x_2^2/4}= e^{- \alpha P_v x_2^2/2}.
\end{equation}
This condition also fixes the spectrum and the splitting constant $\Delta$. Considering $n_i\in \mathbb{N}$ we get from Eq. (\ref{eq:splitting_constants}) 
\begin{equation}
    P_u= \alpha \left[ i \left( n_1+\frac 12\right)- \left( n_2+\frac 12\right)\right], 
    \end{equation}
    and 
    \begin{equation}
    \Delta= \alpha P_v\left[ i \left( n_1+\frac 12\right)+ \left( n_2+\frac 12\right)\right].
\end{equation}
\\

It is now possible to target the non-linear version of the equation in presence of a source
\begin{equation}
     (2\partial_u\partial_v - \alpha^2(x_1^2-x_2^2)\partial_v^2 + \partial_1^2+\partial_2^2)\Phi =\beta\, e^{2i P_v v+ 2i P_u u + i \alpha P_v (x_1^2 + i x_2^2)}\,.
\end{equation}
First we substitute as before a test solution of the form
\begin{equation}
    \Phi =\, e^{2i P_u u + 2i P_v v}\phi_2(x_1)\phi_2(x_2)
\end{equation}
to get
\begin{equation}
    -8 P_u P_v\,\phi_1\phi_2 + 4 \alpha^2 P_v^2 (x_1^2 -x_2^2)\,\phi_1\phi_2+ \phi_2\partial_1^2\phi_1+ \phi_1\partial_2^2\phi_2= \beta\,e^{ i \alpha P_v (x_1^2 + i x_2^2)}. \label{eq:f1f2}
\end{equation}
By defining 
\begin{equation}
    2P_v= \tp,
\end{equation}
Eq. (\ref{eq:f1f2}) is written as 
\begin{equation}
    -4P_u \tp \,\phi_1\phi_2+ \alpha^2 \tp^2 (x_1^2-x_2^2)\,\phi_1\phi_2 + \phi_2\partial_1^2\phi_1+ \phi_1\partial_2^2\phi_2=\beta\, e^{ i \alpha \tp (x_1^2 + i x_2^2)/2},
\end{equation}
where its left-hand side is the same as the linear one.
It is therefore useful to change variables again like in Eq. (\ref{eq:variable_change_to_ho}) but this time with $P_v\to \tp$, to get
\begin{equation}
    -2\frac{P_u}{\alpha} \phi_1\phi_2 -i \phi_2 \hm_1\phi_1 + \phi_1 \hm_2\phi_2 = \frac{\beta}{2\tp \alpha} e^{-y_1^2/4}e^{-y_2^2/4}. 
\end{equation}
It is easy to see that a possible solution of the equation above is 
\begin{equation}
   \phi_1\phi_2 = \tilde{b}_{\Phi}\, e^{-y_1^2/4}e^{-y_2^2/4}
\end{equation}
leading to
\begin{equation}
    \left(-2\frac{P_u}{\alpha}  +\frac i2 -  \frac 12\right)\tilde{b}_{\Phi}\,e^{-y_1^2/4}e^{-y_2^2/4} = \frac{\beta}{2\tp \alpha} e^{-y_1^2/4}e^{-y_2^2/4},
\end{equation}
which gives, solving for $\tilde{b}_{\Phi}$
\begin{equation}
    \tilde{b}_{\Phi}=\frac{\beta\, e^{i\pi/4}}{\sqrt{2}\tp \alpha}.
\end{equation}
We see that the second-order equation with source the square of the linear solution is solved by the source itself, up to a constant that takes into account the mismatch in the ``ground state" energy.
In other words, the solution of the non-homogeneous equation for the second-order perturbation, is the square of the homogeneous first order solution, up to constant.
Furthermore the non-linear solution can be written in the exact same form as the linear counterpart by performing the change of variables with $\widetilde{P}_v$.
This means that if we match the linear solution with the WKB mode, it is possible to do the same also with the non-linear one, given that both the solutions solve the equation $H\psi=-\psi/2$ and the connection with the WKB is unique in the case of an harmonic oscillator.

We now proceed with the matching of a generic solution with its asymptotic WKB approximation.
Assuming the differential equation for the harmonic oscillator holds $y\in (-\infty,\infty)$, we have that the solution is approximately
\begin{equation}
    \psi_1=\frac{C}{(-\bar Q^{1/4}(y))}\exp\left(-\int_y^{-a} \sqrt{-\bar Q(y')}
    \dd y'\right),\quad y<-a,
\end{equation}

\begin{equation}
    \psi_2=\frac{2C}{\bar Q^{1/4}(y)}\cos\left(\int_{-a}^y \sqrt{\bar  Q(y')}
    \dd y'-\frac{\pi}{4}\right),\quad -a<y<a,
\end{equation}

\begin{equation}
    \psi_3=\frac{C}{(-\bar Q^{1/4}(y))}\exp\left(-\int_{a}^y \sqrt{-\bar Q(y')}
    \dd y'\right),\quad a<y,
\end{equation}
where 
\begin{equation}
    \bar Q(y)=-\frac 14 (y^2 -a^2). 
\end{equation}
In addition, the inversion point is at
\begin{equation}
    a=2\sqrt{n + \frac 12},
\end{equation}
and $C$ is the matching constant.
All the integrals can be performed exactly, leading to
\begin{align}
\label{eq:wkb_quadratic_approx}
    \psi_1=&\frac{C}{\left|\frac 14 (y^2 -a^2)\right|^{1/4}}\exp\left(\frac{1}{4} y \sqrt{y^2-a^2}-\frac{a^2}{4} \log\left(-\frac{y+\sqrt{y^2-a^2}}{a}\right)\right),\quad y<-a,\nonumber\\
    \psi_2=&\frac{2C}{\left|\frac 14 (y^2 -a^2)\right|^{1/4}}\cos\left(\frac{1}{4} \left(y \sqrt{a^2-y^2}+a^2 \sin ^{-1}\left(\frac{y}{a}\right)\right)+\frac{\pi  a^2}{8}-\frac{\pi }{4}\right),\quad -a<y<a,\nonumber\\
    \psi_3=&\frac{C}{\left|\frac 14 (y^2 -a^2)\right|^{1/4}}\exp\left(-\frac{1}{4} y \sqrt{y^2-a^2}-\frac{a^2}{4} \log\left(\frac{y-\sqrt{y^2-a^2}}{a}\right)\right),\quad y>a.\nonumber\\
    &
\end{align}
The matching now between the $\psi_2$ and $\Phi$ in Eq. (\ref{eq:Phi_solution}) at $r=r_0$ (which corresponds to $x_1=x_2=0$), leads to 
\begin{equation}
    C=\frac{\sqrt{\pi } 2^{\frac{n}{2}-\frac{1}{4}} \sqrt[4]{2 n+1} \sec \left(\frac{\pi  \
n}{2}\right)}{2\Gamma \left(\frac{1}{2}-\frac{n}{2}\right)}\,b_\Phi,
\end{equation}
for even values of $n$. In particular for $n=0$ we get
\begin{equation}
    C=\frac{1}{2^{5/4}}\,b_\Phi.
\end{equation}

Now that we have connected the three parts of the solution for the WKB of an harmonic oscillator, we need to perform the matching for the linear and non-linear asymptotic solution. 
We start from the former.
We need to change variables back to the radial coordinate $r$, and use the quadratic approximation for the QNMs potential. 
The latter, for the eikonal limit at large $\ell$ is given by
\begin{equation}
\label{eq:Q_large_l}
    Q(r)\simeq \omega^2_{\ell}- V(r), \quad V(r)=\left(1-\frac{r_s}{r}\right)\frac{\ell^2}{r^2}, 
\end{equation}
where $r_s=2M$.
The equation of motions for a perturbation with this potential has the form
\begin{equation}
\label{eq:general_Q}
    \partial_{r_*}^2\psi + Q(r(r_*))\psi =0,
\end{equation}
with 
\begin{equation}
    r_*=r+r_s \log\left(\frac{r}{rs}-1\right), \quad \frac{\partial r}{\partial r_*}= 1-\frac{r_s}{r}.
\end{equation}
When plugged into the WKB approximation at large $r\sim r_*$, we get the correct asymptotic behavior   
\begin{equation}
\label{eq:asymptotic_solution}
    \psi_{\infty}=\frac{a_3 }{(-Q^{1/4}(r_*))}e^{i \int^{r_*} \sqrt{Q(r')}\dd r'}\sim e^{i\omega r}
\end{equation}
However to perform the matching we are interested at what happens to the function $Q(r)$ close to the peak of the potential.
Close to the peak of the potential, we can approximate $Q(r)$ as before
\begin{equation}
    Q(r_*) \simeq Q_0 + \frac 12 Q''_0\, \delta r_*^2, \qquad \delta r_*= r_* - r_{0*},
\end{equation}
and we can express Eq. (\ref{eq:asymptotic_solution}) in terms of the new variable $y$, defined as
\begin{equation}
    y= \gamma_{\ell}\, \delta r_*, \quad \gamma_{\ell} = e^{-i \pi/4} \left(2 Q_0''\right)^{1/4}
\end{equation}
where, as in Ref \cite{Schutz:1985km}
\begin{equation}
    \frac{Q_0}{\gamma_{\ell}^2}=\left( \frac 12 + n\right).
\end{equation}
By using Eq. (\ref{eq:Q_large_l}), we find that
\begin{equation}
   Q_0=\omega_{\ell}^2-\frac{4 \,\ell^2}{27\,r_s^2},\qquad Q''_0= \partial_{r_*}^2 Q|_{r_{0*}} = \frac{32\,\ell^2}{729\, r_s^4},
\end{equation}
so that Eq. (\ref{eq:general_Q}) turns out to be
\begin{equation}
    \psi''+ \omega_{\ell}^2\psi - \frac{4 \,\ell^2}{27\,r_s^2}\psi + \frac{1}{2} \frac{32\,\ell^2}{729\, r_s^4}(r_*-r_{0*})^2\psi=0.
\end{equation}
Solving this equation and imposing the boundary conditions to obtain the QNMs as in Ref. \cite{Fransen:2023eqj}, we get 
\begin{equation}
    \omega_{\ell}^2= \frac{4 \,\ell^2}{27\,r_s^2}- i\left(n+\frac 12\right)\sqrt{\frac{64\,\ell^2}{729\, r_s^4}}.
\end{equation}
In the large-$\ell$ limit we get the QNM frequencies  
\begin{equation}
    \omega_{\ell} = \frac{2\ell}{3\sqrt{3}r_s} - i \left(n+\frac 12\right)\frac{2}{3\sqrt{3}r_s}\equiv \ell\,\omega - i \left(n+\frac 12\right) \omega,
\end{equation}
where
\begin{equation}
    \omega=\frac{2}{3\sqrt{3}r_s}.
\end{equation}
Substituting back we get
\begin{equation}
     \psi''+ \left(\frac{4\,\ell}{27\, r_s^2}\right)^2(r_*-r_{0*})^2\psi= 2i\left(n+\frac 12\right)\frac{4\,\ell}{27\, r_s^2}\psi,
\end{equation}
which is the same equation we get in the Penrose limit after substituting
\begin{equation}
       3\,(r-r_0)=\sqrt{3}x_1= r_*-r_{0*}.
\end{equation}
Therefore we showed that the equation for the eikonal limit solved in WKB approximation when expanded close to the peak gives the same result as the one in Penrose limit.\\

To complete the matching, the asymptotic solution (\ref{eq:asymptotic_solution}) turns out to be
\begin{equation}
\psi_{\infty}\simeq\frac{a_3\,e^{i\pi/4}}{\sqrt{\gamma_{\ell}}\left(n+\frac 12 - \frac 14 y^2\right)^{1/4}}\exp\left( i\int^{y}_{a} \sqrt{\left(n+\frac 12 - \frac 14 y'{}^2\right)}\dd y'\right),
\end{equation}
which subsequently can then be matched to the solution in Eq. (\ref{eq:wkb_quadratic_approx}), written as 
\begin{eqnarray}
    \psi_{3}&=&\frac{b_{\Phi}}{2^{5/4}(-\bar{Q}(y))^{1/4}}\exp\left( i\int^{y}_{a} \sqrt{\left(n+\frac 12 - \frac 14 y'{}^2\right)}\dd y'\right) \nonumber\\
    &=& \frac{a_3 \,e^{i\pi/4}}{\sqrt{\gamma_{\ell}}\left(n+\frac 12 - \frac 14 z^2\right)^{1/4}}\exp\left( i\int^{z}_{a} \sqrt{\left(n+\frac 12 - \frac 14 z'{}^2\right)}\dd z'\right). 
\end{eqnarray}
The matching specifies the constant $a_3$ to be  
\begin{equation}
    a_3= \frac{\sqrt{\gamma_{\ell}}e^{-i\pi/4}}{2^{5/4}}b_{\Phi}.
\end{equation}
Putting everything together, we have for the linear case
\begin{align}
    \Phi_{\ell,0,0}= &b_\Phi\exp\left(-i\omega_{\ell,0,0}t + i \ell \phi + \frac{9}{2} i \,\ell \omega_{\rm orb} \lambda_L\,  (r-r_0)^2  \right) \to\nonumber\\
    &\frac{\sqrt{\gamma_{\ell}} \,b_\Phi}{2^{5/4} c_{\ell}}\frac{1}{(-Q^{1/4}(r_*))}\exp\left(\int^{r_*}\sqrt{-Q(r')}\dd r'\right)
\end{align}
which then goes to infinity like
\begin{equation}
    \psi_{\infty}\simeq b_\Phi\, \frac{\sqrt{\gamma_{\ell}}}{2^{5/4} \sqrt{\omega_{\ell}}} \frac{1}{c_{\ell}} e^{i\omega (r-t)}{}_{-2}Y_{\ell,\ell}\left(\theta,\phi\right).
\end{equation}
Finally to normalise correctly $\psi_{\infty}$ with respect to the boundary condition $\sim e^{i\omega r}$ we need to rescale $a_3$, which means rescaling
\begin{equation}
    b_{\Phi}\to b_{\Phi} \sqrt{\omega_{\ell}},
\end{equation}
to get
\begin{equation}
\label{eq:asymptotic_renormalized}
    \psi_{\infty}\simeq b_\Phi\, \frac{\sqrt{\gamma_{\ell}}}{2^{5/4}} \frac{1}{c_{\ell}} e^{i\omega (r-t)}{}_{-2}Y_{\ell,\ell}\left(\theta,\phi\right).
\end{equation}
We focus now on the matching for the non-linear solution. This can be performed following almost exactly the same steps as the previous part, however there is the further assumption that the source is localized exactly in the Penrose limit region and the wave propagates freely after that. 
This assumption, albeit strong, is in line with the rest of the literature, given that the asymptotic solution from which the non-linear ratio is extracted is written as a free plane wave. 
Nonetheless we are able to reproduce the result for the non-linear ratio analytically and most of its properties, as shown in section \ref{sec:quadratic_qnm}.

For the non-linear matching we start from
\begin{equation}
    \Phi_{2\ell,0,0}= \widetilde b_\Phi \,\exp\left[2 i P_u u + 2i P_v v + i \alpha \, P_v (x_1+i x_2) \right],
\end{equation}
which is then rewritten as
\begin{equation}
    \Phi_{2\ell,0,0}=  \,\exp\left[2 i P_u u + i \tp v + i \alpha \, \tp (x_1+i x_2)/2 \right],
\end{equation}
or, in the original variables
\begin{equation}
     \Phi^{(2)}_{2\ell,0,0}= \widetilde b_\Phi \,\exp\left[-2i\omega_{\ell,0,0} t +  2i\ell \phi + \frac{9}{2}  \,2i\ell \omega^2\,  (r-r_0)^2 - \frac{27}{2}  2\ell \omega^2\,M^2(\theta-\pi/2)^2 \right].
\end{equation}
This function must be matched both on the angular level and on the radial level. 
We start from the angular part, which it is simply the coefficient $c_{\ell}$ multiplied by a factor of 2,
\begin{equation}
     \widetilde b_\Phi e^{i2\ell \phi} \to \frac{\widetilde b_\Phi}{c_{2\ell}}{}_{-2}Y_{2\ell 2\ell}\left(\frac{\pi}{2},\phi\right).
\end{equation}
For the radial case instead we have
\begin{equation}
    Q^{(2)}(r)= 4\omega_{\ell}^2 - \left(1-\frac{r_s}{r}\right)\frac{4\ell^2}{r^2}\simeq \omega_{2\ell}^2 - \left(1-\frac{r_s}{r}\right)\frac{(2\ell)^2}{r^2},
\end{equation}
where we approximated the frequency
\begin{equation}
    2\omega_{\ell}\simeq \omega_{2\ell},
\end{equation}
because we are taking the large $\ell$ limit and the mismatch in the imaginary part will go to zero like $1/\ell$.
Therefore, it can be seen that the change of variables
\begin{equation}
    z= \widetilde \gamma x = \sqrt{2}\gamma_{\ell} x
\end{equation}
gives the correct mapping. After correctly normalizing as in Eq. (\ref{eq:asymptotic_renormalized}), the result for the non-linear solution is
\begin{equation}
     \psi_{\infty}^{(2)}\simeq \widetilde{b}_\Phi\, \frac{\sqrt{\gamma_{\ell}}}{2} \frac{1}{c_{2\ell}} e^{i2\omega_{\ell} (r-t)}{}_{-2}Y_{2\ell 2\ell}\left(\theta,\phi\right).
\end{equation}
Finally $\widetilde b_\Phi$ is the coefficient in front of the non-linear solution, it is completely fixed, and equals to 
\begin{equation}
    \widetilde b_\Phi = \frac{b_\Phi^2}{4} \frac{\alpha \tp e^{i \pi/4}}{\sqrt{2}\alpha \tp}=b_\Phi^2\frac{e^{i \pi/4}}{4\sqrt{2}}.
\end{equation}
Putting everything together we get that the final result for $\mathcal{R}_{\ell\times \ell}$ in the large $\ell$ case is
\begin{equation}
    \mathcal{R}_{\ell\times \ell} = \frac{1}{2\sqrt{\gamma_{\ell}}} \cdot\frac{c_{\ell}^2}{c_{2\ell}}\frac{h^{(2)}}{\big(h^{(1)}\big)^2}.
\end{equation}

\section{ quadratic QNMs for the channel $(2\times\ell)\to \ell+2$}
\label{app:quadratic_qnm}
\renewcommand{\theequation}{B\arabic{equation}}
\noindent
According to the symmetry argument spelled out in subsection \ref{sec:symmetry_argument}, given a solution for the Hertz potential one can construct a new solution for it by a simple rescaling of the coordinates  (\ref{change}) \cite{Kehagias:2024sgh}.
Under such a transformation, the linearly perturbed metric (\ref{eq:metricp1}) transforms conformally   as ${\rm d} s^2\to \lambda^2\,{\rm d}s^2$ as long as we assign a conformal weight to the Hertz potential equal to four. 
Consider a linear mode with $\ell\gg 2$ and the channel $(2\times\ell)\to \ell+2$. 
Choosing $\lambda=1/\ell^{1/2}\ll 1$, the dependence on the $v$ and ($z,\bar z)$ coordinates in the wave-function (\ref{source2}) satisfying Eq. (\ref{eq:hertz_potential}) is suppressed for the $\ell=2$ mode by $1/\ell$ and once can make use of a residual gauge transformation in the radiation gauge to eliminate such a mode and go back to the background. 
The very same transformation generates a quadratic QNM for $(\ell+2)$ with amplitude \cite{Kehagias:2024sgh}

\be
\Phi^{(2)}_{\ell+2}=\Phi^{(1)}_{\ell}\Phi^{(1)}_2\frac{\omega\ell}{\frac{3}{2}\omega(i-1)}\frac{3}{2}\omega^2(1+i)=-i\omega^2\ell,
\ee
from which we deduce, using Eq. (\ref{comp})

\be
\frac{h^{(2)}_{\ell+2}}{h^{(1)}_{\ell}h^{(1)}_{\ell}}=i\frac{(\ell+2)^2}{4\ell}.
\ee
Performing the matching to go from the Penrose limit to large distances away from the photon ring, assuming that the eikonal  approximation holds also for $\ell=2$, we finally find

\begin{equation}
    |\mathcal{R}_{2\times \ell}|= \frac{(\ell+2)^2}{2^{5/4}\ell}\left|\frac{\sqrt{\gamma_{\ell+2}}}{\sqrt{\gamma_{\ell}\gamma_{2}}} \cdot\frac{c_2 c_{\ell}}{c_{\ell+2}}\right|\underset{\ell\gg 2}{\simeq} 0.11\,\ell,
\end{equation}
as predicted  numerically \cite{Bucciotti:2025rxa}. 
The non-linear effects can be quite accurately and analytically captured through the Penrose limit in the eikonal limit.  
This also shows that QNMs can be interpreted as adiabatic modes \cite{Kehagias:2024sgh}, contrary to what was claimed in Ref. \cite{Bucciotti:2025rxa} where the passage from the Hertz potential to the gravitational waves and the matching procedure have not been performed and it was incorrectly stated that the Penrose limit produces a non-linearity which depends upon the space coordinates\footnote{In Ref. \cite{Bucciotti:2025rxa} it has been also incorrectly stated that the linear adiabatic mode for the gravitational wave does no satisfy the equation of motion. Indeed, it does once     the full space dependence is accounted for  in the equation and Taylor expand around the photon ring.  }.

\section{the issue of gauge invariance at second-order}
\label{app:gauge_invariance}
\renewcommand{\theequation}{C\arabic{equation}}
\noindent
Consider a coordinate transformation of the type (active transformation in some physical point $P$)
\begin{equation}
\widetilde{x}^\mu(P)  = x^\mu(P)+\epsilon\,\xi_1^\mu(P) +\frac{\epsilon^2}{2} \left[\partial_\nu\xi_{1}^\mu(P)\xi_1^\nu(P)+\xi_2^\mu(P)\right].
\end{equation}
Consider now the scalar $\Psi_0$ for which
\begin{eqnarray}
\Psi_0(x^\mu)&=&\Psi_0^{(0)}(x^\mu)+\epsilon\,\Psi_0^{(0)}(x^\mu) +\frac{\epsilon^2}{2}  \Psi_0^{(2)}(x^\mu),\nonumber\\
\widetilde\Psi_0(\widetilde x^\mu)&=&\Psi_0^{(0)}(\widetilde x^\mu)+\epsilon\,\widetilde\Psi_0^{(1)}(\widetilde x^\mu) +\frac{\epsilon^2}{2}  \widetilde\Psi_0^{(2)}(\widetilde x^\mu)\nonumber\\
&=&\Psi_0^{(0)}\left(x^\mu+\epsilon\,\xi_1^\mu +\frac{\epsilon^2}{2} \partial_\nu\xi_{1}^\mu\xi_1^\nu+\frac{\epsilon^2}{2} \xi_2^\mu\right)\nonumber\\
&+&\epsilon\,\widetilde\Psi_0^{(0)}(x^\mu+\epsilon\,\xi_1^\mu)\nonumber\\
&+&\frac{\epsilon^2}{2}  \widetilde\Psi_0^{(2)}(x^\mu)\nonumber\\
&=&\Psi_0^{(0)}( x^\mu)+\epsilon\left[\widetilde\Psi_0^{(0)}(x^\mu)+\partial_\nu\Psi_0^{(0)}(x^\mu)\xi_1^\nu\right]\nonumber\\
&+&\frac{\epsilon^2}{2}\left[\widetilde\Psi_0^{(2)}(x^\mu)+
\partial_\nu\Psi_0^{(0)}(x^\mu)\xi_2^\nu+
\partial_\sigma\partial_\nu\Psi_0^{(0)}(x^\mu)\xi_1^\sigma\xi_1^\nu+
\partial_\sigma\Psi_0^{(0)}(x^\mu)\partial_\nu\xi_{1}^\sigma\xi_1^\nu+2\partial_\nu\Psi_0^{(1)}(x^\mu)\xi_1^\nu 
\right].\nonumber\\
&&
\end{eqnarray}
Since $\Psi_0$ is a scalar, we have 
\begin{equation}
\widetilde\Psi_0(\widetilde{x})=\Psi_0(x),
\end{equation}
and we finally obtain
\begin{eqnarray}
   \widetilde\Psi_0^{(1)}( x)  &=&\Psi_0^{(1)}( x)-\partial_\mu\Psi_0^{(0)}( x)\xi_1^\mu,\nonumber\\
   \widetilde\Psi_0^{(2)}( x)  &=&\Psi_0^{(2)}( x)-\partial_\mu\Psi_0^{(0)}( x)\xi_2^\mu
   -
\partial_\sigma\partial_\nu\Psi_0^{(0)}(x^\mu)\xi_1^\sigma\xi_1^\nu-
\partial_\sigma\Psi_0^{(0)}(x^\mu)\partial_\nu\xi_{1}^\sigma\xi_1^\nu-2\partial_\nu\Psi_0^{(1)}(x^\mu)\xi_1^\nu.\nonumber\\
&&
\end{eqnarray}
The background value of $\Psi_0$ is zero, therefore
\begin{eqnarray}
   \widetilde\Psi_0^{(1)}( x)  &=&\Psi_0^{(1)}( x),\nonumber\\
   \widetilde\Psi_0^{(2)}( x)  &=&\Psi_0^{(2)}( x)-2\partial_0\Psi_0^{(1)}(x)\xi_1^0-2\partial_i\Psi_0^{(1)}(x)\xi_1^i,
\end{eqnarray}
showing that at first-order the perturbation is gauge-invariant, but at second-order is not. This is the same for $\Psi_4$, it is gauge-invariant at first-order as $\Psi_4^{(0)}=H_{\bar z\bar z}/2=1/(3M^2)$ is constant; at second-order though gauge invariance is lost. Therefore, to compare to numerical results, done in the (outgoing) radiation gauge, it is crucial to adopt the very same gauge at second-order.

\newpage
\bibliography{draft.bib}

\end{document}